\documentclass{emulateapj}
\usepackage{appendix}

\def\cm3{\mbox{cm}^{-3}}

\def\tdec2{t_{dec,2}}
\def\cm{\mbox{cm}}

\slugcomment{ }

\begin{document}
\title{Inefficient cosmic ray diffusion around Vela X : constraints from H.E.S.S. observations of very high-energy electrons}

\author{Zhi-Qiu Huang\altaffilmark{1,2}, Kun Fang\altaffilmark{4,5}, Ruo-Yu Liu\altaffilmark{3}, Xiang-Yu Wang\altaffilmark{1,2}}

\altaffiltext{1}{School of Astronomy and Space Science, Nanjing University, Nanjing 210093, China;
xywang@nju.edu.cn}
\altaffiltext{2}{Key laboratory of Modern Astronomy and Astrophysics (Nanjing University), Ministry of Education, Nanjing 210093, China}
\altaffiltext{3}{Deutsches Elektronen Synchrotron (DESY), Platanenallee 6, D-15738 Zeuthen, Germany}
\altaffiltext{4}{Key Laboratory of Particle Astrophysics, Institute of High Energy Physics, Chinese Academy of
Sciences, Beijing 100049, China}

\altaffiltext{5}{School of Physics, University of Chinese Academy of Sciences, Beijing 100049, China}

\begin{abstract}
Vela X is a nearby pulsar wind nebula (PWN) powered by a $\sim 10^4$ year old pulsar. Modeling of the spectral energy distribution of the Vela X PWN has shown that accelerated electrons  have largely escaped from the confinement, which is likely due to the disruption of the initially confined PWN by the SNR reverse shock. The escaped  electrons  propagate to the earth and contribute to the measured local cosmic-ray (CR) electron spectrum. We find that the escaped CR electrons from Vela X would hugely exceed the measured flux by HESS at $\sim 10$ TeV  if the standard diffusion coefficient for the interstellar medium is used.
We propose that the diffusion may be highly inefficient around Vela X and find that a spatially-dependent diffusion can lead to CR flux consistent with the HESS measurement. Using a two-zone model for the diffusion around Vela X, we find that  the diffusion coefficient in the inner region of a few tens of parsecs should be  $\la10^{28}{\rm cm^2 s^{-1}}$ for $\sim10$ TeV CR electrons, which is about two orders of magnitude lower than the standard value for ISM.
Such inefficient diffusion around PWN resembles the case of the Geminga and Monogem PWNe, suggesting that inefficient diffusion  may be common in the vicinity of PWNe spanning a wide range of ages.
\end{abstract}


%%%%%%%%%%%%%%%%%%%%%%%%%%%%%%%%%%%%%%%%%%%%%%%%%%%%%%%%%%%%%%%%
\section{Introduction}
Recent measurements of an increasing CR positron ($e^+$) fraction above 10 GeV by PAMELA and AMS-02 identify an excess of high-energy positrons relative to the standard predictions for secondary electron-positron production in the interstellar medium (ISM)\citep{2010PhRvL.105l1101A,2013PhRvL.110n1102A}, suggesting that these positrons are produced as primary particles. The CR $e^++e^-$  spectrum has been measured up  to a few TeV by Fermi-LAT, Veritas, DAMPE \citep{2017PhRvD..95h2007A,2015ICRC...34..411S,2017Natur.552...63D} and to $\sim 20$ TeV by HESS recently \citep{2017arXiv170906442H}.  As high-energy electrons (hereafter we do not distinguish positrons from electrons) cool efficiently in the ISM through synchrotron and inverse-Compton radiation, they cannot travel beyond a kiloparsec distance before depleting their energies. Therefore, these high-energy electrons must be produced by nearby sources, such as pulsars (or PWNe), annihilating dark matter particles and supernova remnants. While the real sources are still under debate, nearby pulsars such as Geminga, PSR B0656+14 and Vela, which are at distances of only $\sim200$ pc, are the most attractive candidates \citep{1970ApJ...162L.181S,1995PhRvD..52.3265A,2017PhRvD..96j3013H}.

Vela X, powered by a $\sim 10^4$ year old pulsar (i.e., the Vela pulsar or PSR B0833-45), is one of the most well-studied PWN. It consists of an extended radio halo of size $2^\circ \times 3^\circ$ and a small collimated structure, e.g. the "cocoon". High-energy gamma-ray emission has been detected from both the halo and the cocoon by Fermi/LAT and HESS. \citet{2008ApJ...689L.125D} propose that there are two distinct populations of electrons, one responsible for the radio and GeV gamma-ray emission and the other for the X-ray and TeV emission. To explain the steep GeV spectrum measured by Fermi/LAT and the dimness of the TeV nebula relative to the spin-down power of the Vela pulsar, \cite{2011ApJ...743L...7H} argue that significant diffusive escape of electrons from the halo must have occurred. The same conclusion is reached more recently by \cite{2018arXiv180611499T} with an analysis of $\sim 9.5$ years of  data from Fermi/LAT observation.  While the confinement of particles in PWNe is thought to be effective in the early stage, the interaction with the SNR reverse shock, which seems to have appeared in Vela X several thousand years ago \citep{2001ApJ...563..806B,2009ApJ...703.2051G}, may have brought an end to the confinement. The asymmetric structure of the PWN with respect to the pulsar supports such an interpretation. The interaction
is expected to disrupt the PWN sufficiently that diffusion of particles out of the PWN becomes possible.

The escaped electrons will contribute to the CRe spectrum measured at the earth. Using a diffusion coefficient of $1.07\times10^{27} (E/1{\rm GeV})^{0.6} {\rm cm^2 s^{-1}}$ and a total electron energy of $6.8\times10^{48}{\rm erg}$, \cite{2011ApJ...743L...7H} predict a distinct bump in the CR electron spectrum at several TeV. This diffusion coefficient is actually smaller than the standard one for ISM, which is $D_{\rm ISM}\simeq 3.86\times10^{28} {\rm (E_e/GeV)^{0.33} cm^2 s^{-1}}$, as inferred from the boron-to-carbon ratio and other CR secondary-to-primary ratio for electrons with $E\lesssim 10 \rm{TeV}$ (see http://galprop.stanford.edu/). If one uses the standard diffusion coefficient, the bump would be even higher. However, the latest results from HESS do not show such a bump \citep{2017arXiv170906442H}, which invalidates the simple diffusion model for the escape of particles from Vela X.

Interestingly, recent TeV observations of Geminga PWN with the HAWC telescope have been interpreted as evidence that diffusion of high-energy electrons within PWN is highly inefficient compared to the standard value for ISM \citep{2017Sci...358..911A,2017PhRvD..96j3013H}.
Motivated by this discovery, we here study whether a spatially-dependent diffusion around Vela X can lead to a CR electron flux consistent with that measured by HESS. We make a two-zone approximation for the diffusion coefficient surrounding Vela X, i.e., an inner inefficient diffusion zone and an outer standard diffusion zone. We present an analytic approach to calculate the CR electron flux at earth. Although this analytic approach is a crude approximation, it provides a convenient way to estimate the CR flux for the two-zone model. In \S 2, we first present the difficulties for the simple one-zone model with a standard diffusion coefficient. Then, we study the two-zone model and obtain the constraint on the diffusion coefficient of the inner zone around Vela X in \S 3. Finally, we give the discussions and conclusions in \S 4.

\section{Results for  spatially-independent diffusion}
We first study whether a simple one-zone diffusion model, as usually used for CR propagation in ISM, can produce a flux consistent with the HESS measurement.
This simple model assumes that the diffusion is homogeneous along the path from the PWN to the earth. The diffusion coefficient is given by the standard one, $D_{\rm ISM}\simeq 3.86\times10^{28} {\rm (E_e/GeV)^{0.33} cm^2 s^{-1}}$.

The transport of CR electrons can be described by the equation:
\begin{eqnarray}
\frac{\partial}{\partial t}n_e(E_e,\overrightarrow{x},t)&& = \overrightarrow{\nabla} \cdot \left[D(E_{e},\overrightarrow{x})\overrightarrow{\nabla}n_e(E_{e},\overrightarrow{x},t)\right] \\ \nonumber
+ && \frac{\partial}{\partial E_e}\left[\frac{dE_e}{dt}n_e(E_e,\overrightarrow{x},t)\right]
+ Q(E_e,\overrightarrow{x},t),
\end{eqnarray}
where $n_e(E)$ is the differential number density of electrons, $D(E_{e})$ is the diffusion coefficient and $Q(E_e,\overrightarrow{x},t)$ is the source term. Energy losses caused by inverse Compton and synchrotron processes are described as
\begin{eqnarray}
 - \frac{dE_e}{dt}(r) && = \sum  \frac{4}{3}\sigma_T \rho_i(r) S_i(E_e) \left(\frac{E_e}{m_e}\right)^{2} \\ \nonumber
&& + \frac{4}{3} \sigma_T \rho_{mag}(r) \left(\frac{E_e}{m_e}\right)^{2},
\end{eqnarray}
where $\sigma_T$ is the Thomson cross section, $\rho_i$ denotes the radiation energy density of background photons and $\rho_{mag}$ denotes the energy density of the magnetic field. Various components of the radiation backgrounds are taken into consideration, including the cosmic microwave background (CMB), starlight (star), ultraviolet emission (UV) and infrared emission (IR). Parameters are adopted as follows for the area surrounding Vela: $\rho_{\rm{CMB}} =0.260 \rm{eV/cm^{3}}$, $\rho_{\rm{star}} =0.44 \rm{eV/cm^{3}}$, $\rho_{\rm{UV}} =0.10 \rm{eV/cm^{3}}$, $\rho_{\rm{IR}} =0.44 \rm{eV/cm^{3}}$, $\rho_{\rm{mag}} =0.622 \rm{eV/cm^{3}}$ (corresponding to $B=5\rm{\mu G}$) and $T_{\rm{CMB}} = 2.7 \rm{K}$, $T_{\rm{star}} = 7500 \rm{K}$, $T_{\rm{UV}} = 20000 \rm{K}$, and $T_{\rm{IR}} = 25 \rm{K}$ \citep{2008ApJ...689L.125D,2013ApJ...774..110G}. We also check the results adopting other values of these parameters  \citep{2017PhRvD..96j3013H,2018ApJ...863...30F}, but  negligible differences are found.  When $E_e \gtrsim m_e ^{2}/2T$, the suppression of inverse Compton scattering by the Klein-Nishina effect cannot be ignored, which can be parameterized by \citep{Longair2011}
\begin{equation}
S_i(E_e)\approx \frac{45m_{e}^{2}/64\pi ^{2} T_{i} ^{2}}{(45m_{e}^{2}/64\pi ^{2} T_{i} ^{2})+(E_{e}^{2}/m_{e}^{2})}.
\end{equation}

For a burst-like injection of $Q(E_e,t)=\delta (t) Q_0 E_e^{-\alpha}{\rm exp} (-E_e/E_c)$, the solution to Eq.~(1) is given by
\begin{equation}
n_e(E_e, r, t)=\frac{Q_0 E_0 ^{2-\alpha}e^{- E_0/E_c}}{8\pi ^{3/2} E_e ^{2} L_{\rm{dif}}^{3}(E_e,t)}
{\rm{exp}} \left[-\frac{r^2}{4L_{\rm{dif}}^{2}(E_e,t)}\right],
\end{equation}
where $E_0$ is the initial energy of the electron of energy $E_e$ at $t$ and $L_{\rm{dif}}$ is the diffusion length scale given by,
\begin{equation}
L_{\rm{dif}} \equiv \left[\int_{E_0}^{E_e} \frac{D(E^{\prime})}{-dE_e /dt(E^{\prime})} dE^{\prime}\right]^{1/2}.
\end{equation}

We note that Eq.4 permits a fraction of particles to propagate faster than the speed of light, which is the so called superluminal diffusion problem \citep{2007PhRvD..75d3001D,2009ApJ...693.1275A}. From the  mathematical point of view, the superluminal propagation always exists  in the solutions of the non-relativistic diffusion equations, but very frequently the contribution of unphysical regions to the solution is negligibly small. In these cases one can regard that the diffusion equation gives a correct description of the considered physical phenomenon. However, when $t \sim r/c$, the problem of superluminal diffusion in Eq.4 becomes severe. \citet{2009ApJ...693.1275A} find a solution to this problem: if cooling of particles is unimportant (which is applicable to our case), the probability to find one electron at
distance $r$  from the source  at a time $t$ after its injection can be described by
\begin{equation}
P(E_e,t,r)=\frac{\theta (ct-r)}{(ct)^3 Z(\frac{c^2 t}{2D})[1-(\frac{r}{ct})^2]^2} {\rm{exp}}\left[-\frac{\frac{c^2 t}{2D}}{\sqrt{1-(\frac{r}{ct})^2}}\right]
\end{equation}
where
\begin{equation}
Z(y)=4\pi K_1 (y)/y
\end{equation}
with $\theta (r)$ being the Heaviside function and $K_1 (y)$  being the modified Bessel function. Note that the above formula only works when $t$ is much smaller than the cooling timescale of electrons, which is true for our following calculation. The electron density at time $t$ after the injection then can be obtained by
\begin{equation}
n_e(E_e, r, t)=\int_{-\infty}^{t} P(E_e,t -t',r)\delta (t') Q_0 E_e^{-\alpha}e^{-E_e/E_c} dt'
\end{equation}

Following \cite{2011ApJ...743L...7H}, we take $\alpha=1.8$ and $E_c =6$ TeV for the electron spectrum. The total energy injected into the initially confined PWN depends on the birth-period $P_0$ of the pulsar. The energy in relativistic electrons is about $\sim 2\times 10^{49} \epsilon (P_0/30 {\rm ms})^{-2}$, where $\epsilon$ is the fraction of spin-down power converted into relativistic electrons. All previous estimates of the total energy give a value of several $10^{48}{\rm erg}$ \citep{2008ApJ...689L.125D,2010ApJ...714..927A,2011ApJ...743L...7H,2013ApJ...774..110G}. This is consistent with the estimate of the birth-period of $P_0=40$ ms that is invoked to account for the ratio between the PWN radius and  SNR radius \citep{2001ApJ...555L..49V}. In the following calculation, we use a total energy of $6.8\times 10^{48}$ erg, as obtained from the SED fit of Vela X by \citet{2011ApJ...743L...7H}. We use $r=270 \rm{pc}$ as the distance between the earth and Vela.
We assume that the electrons are released instantaneously at the disruption time $t$ of the initial PWN. The exact disruption time of Vela X is unknown, but the reasonable values should be at least several  kyr (e.g., \cite{2001ApJ...563..806B,2009ApJ...703.2051G}).
Note that the time span between the disruption time of Vela X and the current time is exactly the propagation time of injected electrons $t$.

The results of the CR electron spectrum for various $t$  are shown in Fig.1, where Eq.8 is used to avoid possible superluminal diffusion problem\footnote{We find that when $t\ga 2000 \rm{yr}$, the difference between the result of  Eq.4 and Eq.8 can be neglected. Therefore, it is acceptable to use Eq.4 when $t \ga 2000 \rm{yr}$, which also incorporates the influence of electron cooling.}. Obviously, the CR flux produced by Vela X exceed the measured flux by several order of magnitudes for any reasonable value of $t$. Only for a very small $t$ (i.e. $t<1100 \rm{yr}$), the CR flux can be consistent with the HESS measurement (see the blue line in Fig.1). Such a small $t$ seems unreasonable since the timescale when the PWN collides with the SNR  reverse shock  explosion is usually several thousand years (e.g., \cite{2001ApJ...563..806B,2009ApJ...703.2051G}). This suggests that the simple one-zone diffusion model does not work for Vela X.

\begin{figure}
\begin{center}
\includegraphics[scale=0.3]{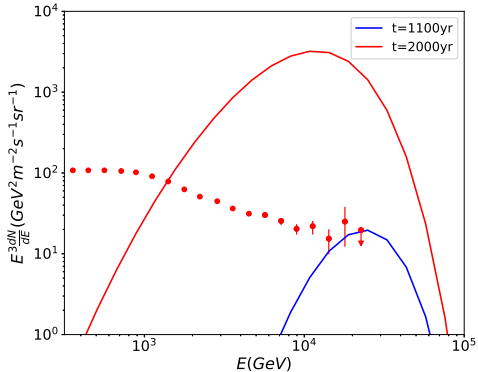}
\caption{The CR electron spectrum produced by Vela X, compared with recent HESS data \citep{2017arXiv170906442H}. The red and blue lines represent the injection times of $t=$ 2kyr and 1.1kyr respectively. A spatially-independent diffusion with a coefficient of $D(E)=3.86 \times 10^{28} (E/\rm{GeV})^{0.33} \rm{cm^{2}/s}$ is adopted.}
\end{center}
\end{figure}

\section{The two-zone diffusion model }
Recent HAWC observations of the PWN regions around Geminga and Monogem provide detailed information about the spatially extended emission of TeV gamma-rays. The spectrum and morphology of the TeV emission can be used to infer the features of underlying high-energy electrons responsible for the IC photons. The angular profiles of the TeV emission observed from Geminga and Monogem indicate that the diffusion is highly inefficient in the regions surrounding these sources \citep{2017Sci...358..911A}. This is also the first empirical determination of a diffusion coefficient in the region of tens of pc around pulsars in the local Galaxy. The inferred diffusion coefficient is more than two orders of magnitude smaller than the standard diffusion coefficient in ISM. Furthermore, assuming a spatially dependent diffusion, \cite{2018ApJ...854...57F} and \cite{2018PhRvD..97l3008P} show that nearby pulsars, such as Geminga,  could  contribute significantly to the CR electron spectrum in $0.1-1$\,TeV. Motivated by these results, we study whether a spatially-dependent diffusion model with  inefficient diffusion in the inner region surrounding Vela X could resolve the inconsistency between the predicted CR electron flux and the observed one by HESS. We approximate the diffusion coefficient of the entire space to be a step function of the distance to Vela X, where the diffusion is suppressed with a coefficient of $D_1$ within a few tens of parsecs from Vela X and a standard diffusion coefficient $D_2$  for the outside region, i.e.
\begin{equation}
D(r)=\{
\begin{array}{ll}
D_1, \,\,\,\,\,\, r\le r_0\\
D_2=D_{\rm ISM}, \,\, r>r_0
\end{array}
\end{equation}

We develop an analytic approach to solve Eq.~(1) in the above two-zone diffusion model. We first use the solution Eq.~(4) with a diffusion coefficient $D_1$ to estimate the number density of electrons at the interface of the two zones (i.e., at the spherical surface with a radius $r_0$). We note that Eq.~(4) is strictly correct only when $D_1=D_2$. So the above obtained number density is a crude estimate when $D_1\neq D_2$. The flux density at the sphere is $F_D = -D\frac{\partial n_e}{\partial r}$ and the number of particles in unit time passing an surface element on the sphere outwardly is $\Delta \dot{N}=F_D\Delta S$, with $\Delta S$ being the area of the element surface. We then regard the surface element as a point source with injection rate $\Delta \dot{N}$. Since the flux at the sphere is a function of time, $\Delta \dot{N}$ is also a function of time.  We then regard the sphere consist of many surface elements, each of which is treated as a point source. A convenient choice is to divide the sphere into annular rings perpendicular to the line connecting the earth and Vela X, so that the distance between the earth and each surface element on the ring is the same. The radius of the ring is $r_0\cdot sin\theta$ and the distance to the earth is $d=(r^2 -2r\cdot r_0 cos\theta + r_0 ^2)^{1/2}$. The area of this ring is $\Delta S_{ring}=2\pi r_0 ^{2} sin\theta d\theta$. The integral can be done over the annular rings on the spherical surface. We further regard the continuous injection from the sphere as the sum of a series of discrete injection so the electron number density at the radius of the earth can be calculated with Eq.~(4). The precision of this analytic method is tested in the appendix. We find that the  difference of the CR electron flux at earth between our analytic results and the numerical approach used by \citet{2018ApJ...863...30F} is at most $\sim 30\%$. Since the uncertainty of the CR flux  measured by HESS at $\sim 10$ TeV is about a factor of two, we consider that the precision of our analytic approach is sufficient for this study.

\begin{figure}
\begin{center}
\includegraphics[scale=0.35]{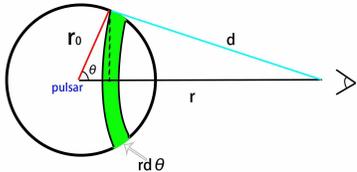}
\caption{A schematic picture of the two-zone diffusion model. $r_0$ is the radius of the region with inefficient diffusion and $r$ is the distance between Vela pulsar and the earth. }
\end{center}
\end{figure}

{{We note that the burst-like injection is an assumption based on the reverse shock interaction scenario. To be more complete, we also consider
the case of continuous injection. Since the injection rate is most likely to be monotonically decreasing, a flat injection profile (i.e., a constant injection rate of electrons) would provide the most conservative estimate  of the electron flux arriving at the Earth, given other parameters are the same. That is,  the burst-like profile and the flat time profile  can be regarded as the two
extremes for any injection patterns.  To deal with the flat injection, we decompose the injection function into many small time bins, treat each time bin as a burst-like injection, and lastly sum up the contribution from each of them. We compare the results of the burst-like injection and flat injection in Fig.~3 for different  time $t$ at which electron injection started.
We find that, in the case of the flat profile injection, a larger diffusion coefficient is obtained. This is due to that a significant part of electrons are injected at later time in this case and these electrons have not arrived at the Earth. The peak of the spectrum  moves towards lower energy for earlier injection  (i.e., larger $t$). This is because that electrons with lower energy can reach the earth while electrons with higher energy will be cooled down. We  find that, for all the considered $t$ here, the difference in the obtained $D_{1}$ for the two different injection profiles is within a factor of 2. We thus conclude that our result is not  sensitive to the injection profile and in what follows we only consider the burst-like injection profile for simplicity.}}

\begin{figure*}
\begin{center}
\includegraphics[scale=0.34]{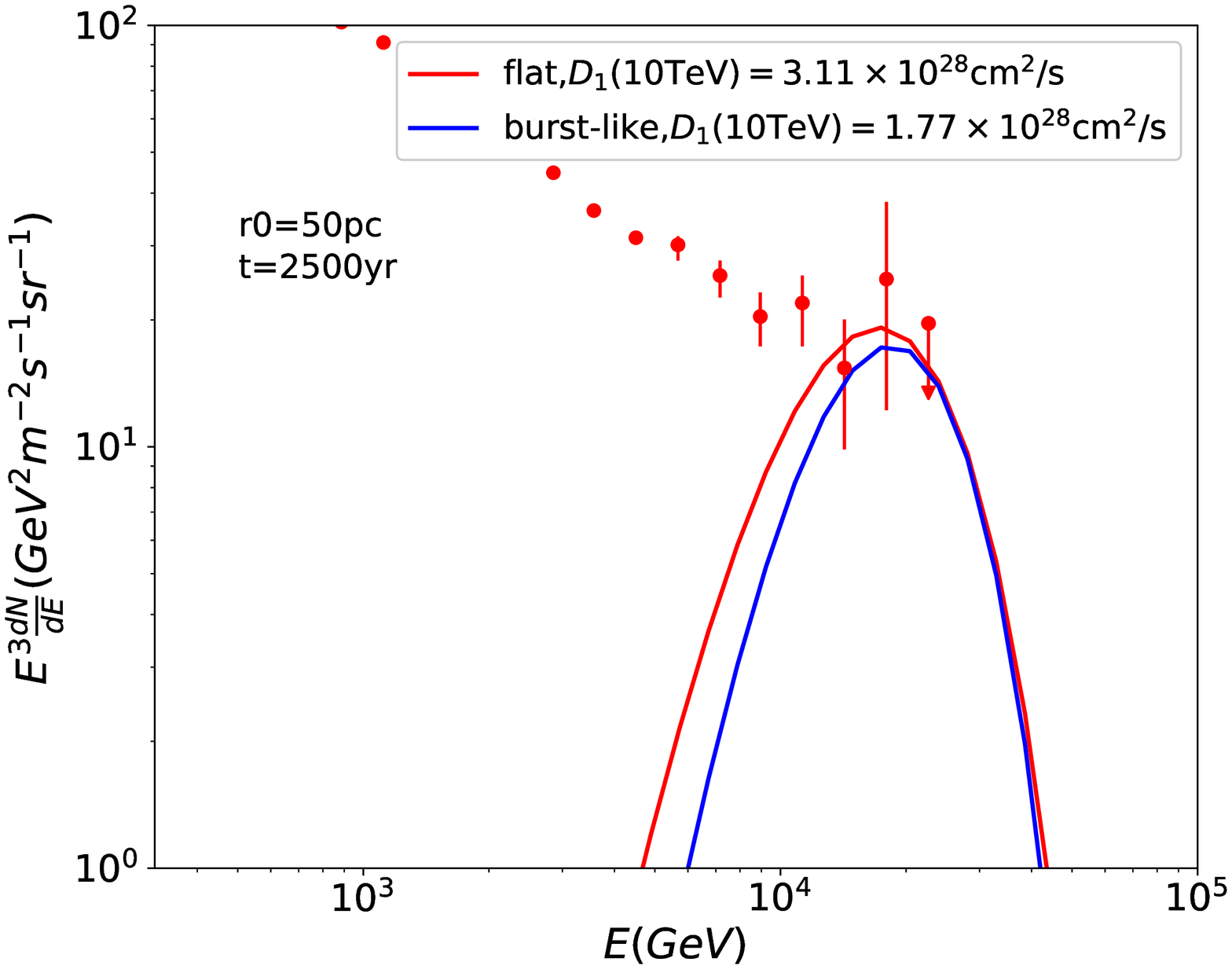}
\includegraphics[scale=0.34]{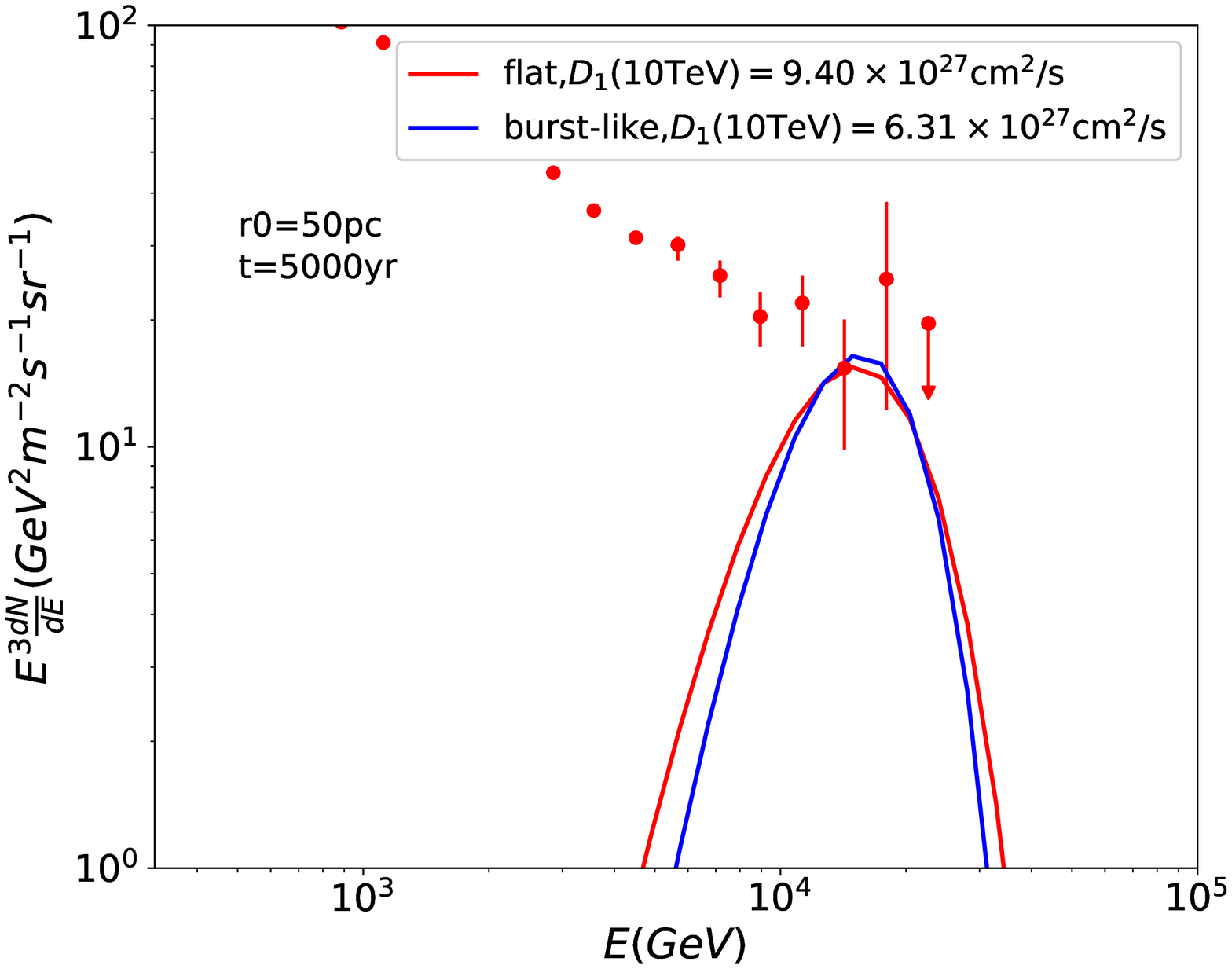}
\includegraphics[scale=0.34]{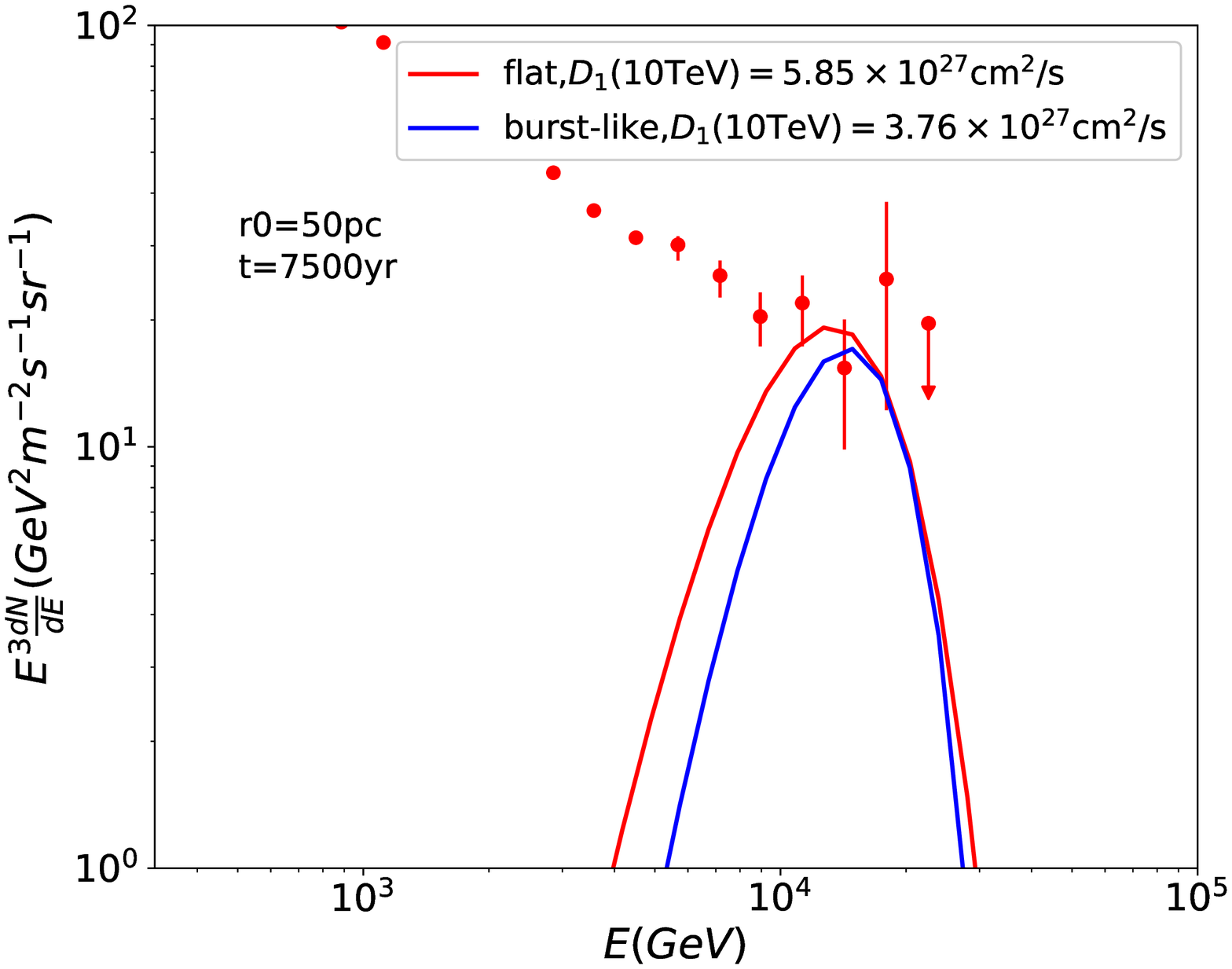}
\includegraphics[scale=0.34]{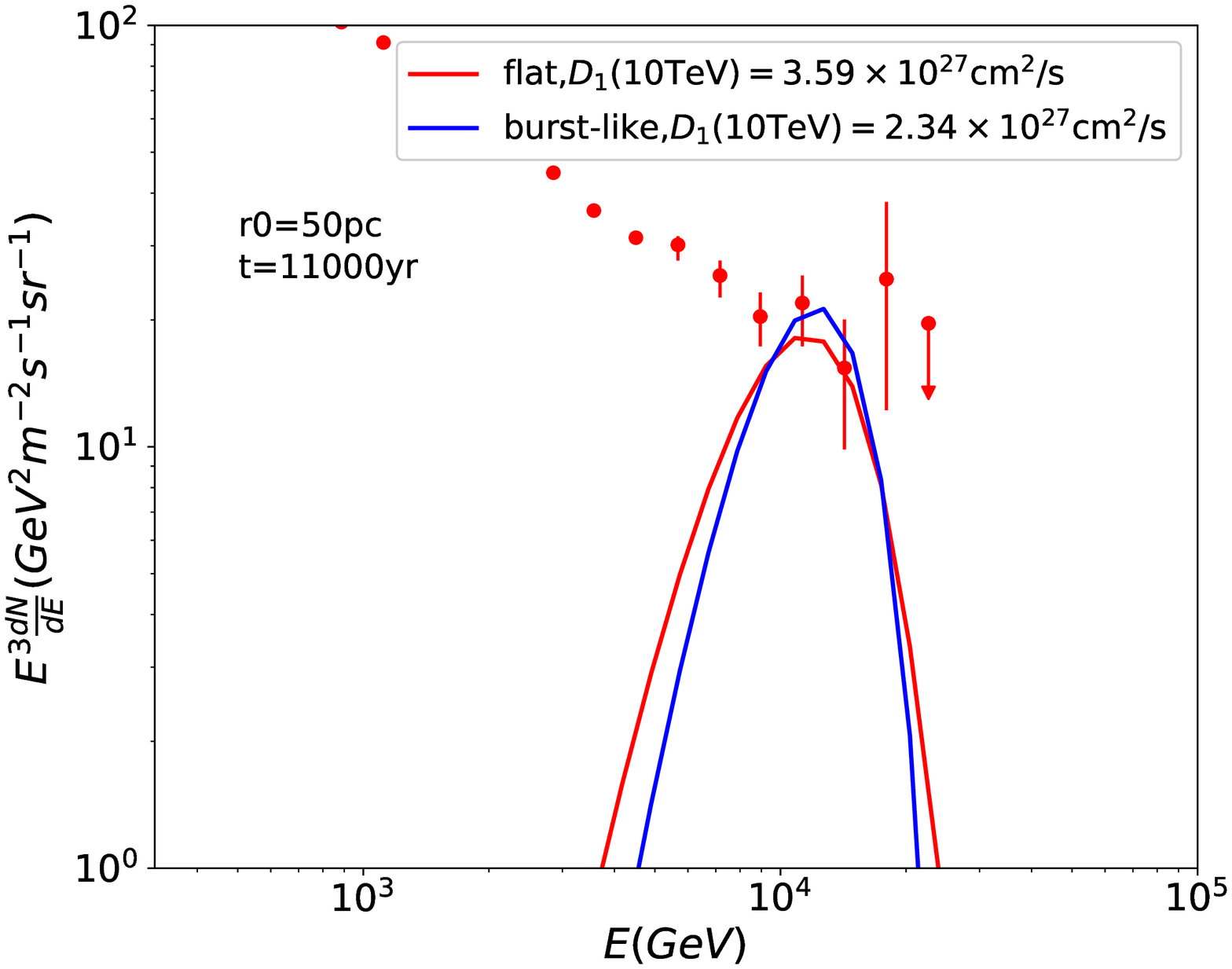}
\caption{Comparison between the CR electron flux produced by  Vela X in the two-zone model and the measured flux by HESS. $t$ is the time when the injection started. $r_0=$ 50 pc is adopted. The red lines show the results for  flat injections while the blue lines show the results for burst-like injections. The diffusion coefficients in both inner and outer zones are assumed to be proportional to $E^{0.33}$ (specifically, $D_2(E)=3.86 \times 10^{28} (E/{\rm{GeV}})^{0.33} \rm{cm^{2}/s}$).}
\end{center}
\end{figure*}

The disruption time $t$ of the PWN and the radius of the inner zone are two unknown parameters. We use  $r_0=30$ pc and $r_0=50$ pc   as two reference values for the radius of the inner zone. The CR fluxes for the two-zone model with  $r_0=30$ pc and $r_0=50$ pc  are, respectively, shown in the left and right panels of Fig.4.  For a reasonable range of  $t$ (i.e., $t>2.5$ kyr), we find that only when $D_1({\rm 10 TeV})\la10^{28}{\rm cm^2 s^{-1}}$, the CR electron flux can be consistent with the HESS measurement. This value is about two orders of magnitude smaller than the standard diffusion coefficient at $\sim10$ TeV, which is $D_{\rm ISM}(\rm{10 TeV})\simeq 0.8\times 10^{30}{\rm cm^2 s^{-1}}$. The constraint on the diffusion coefficient becomes more stringent for a larger $t$. This is because a longer diffusion time requires a smaller diffusion coefficient in the inner zone.

\begin{figure*}
\begin{center}
\includegraphics[scale=0.34]{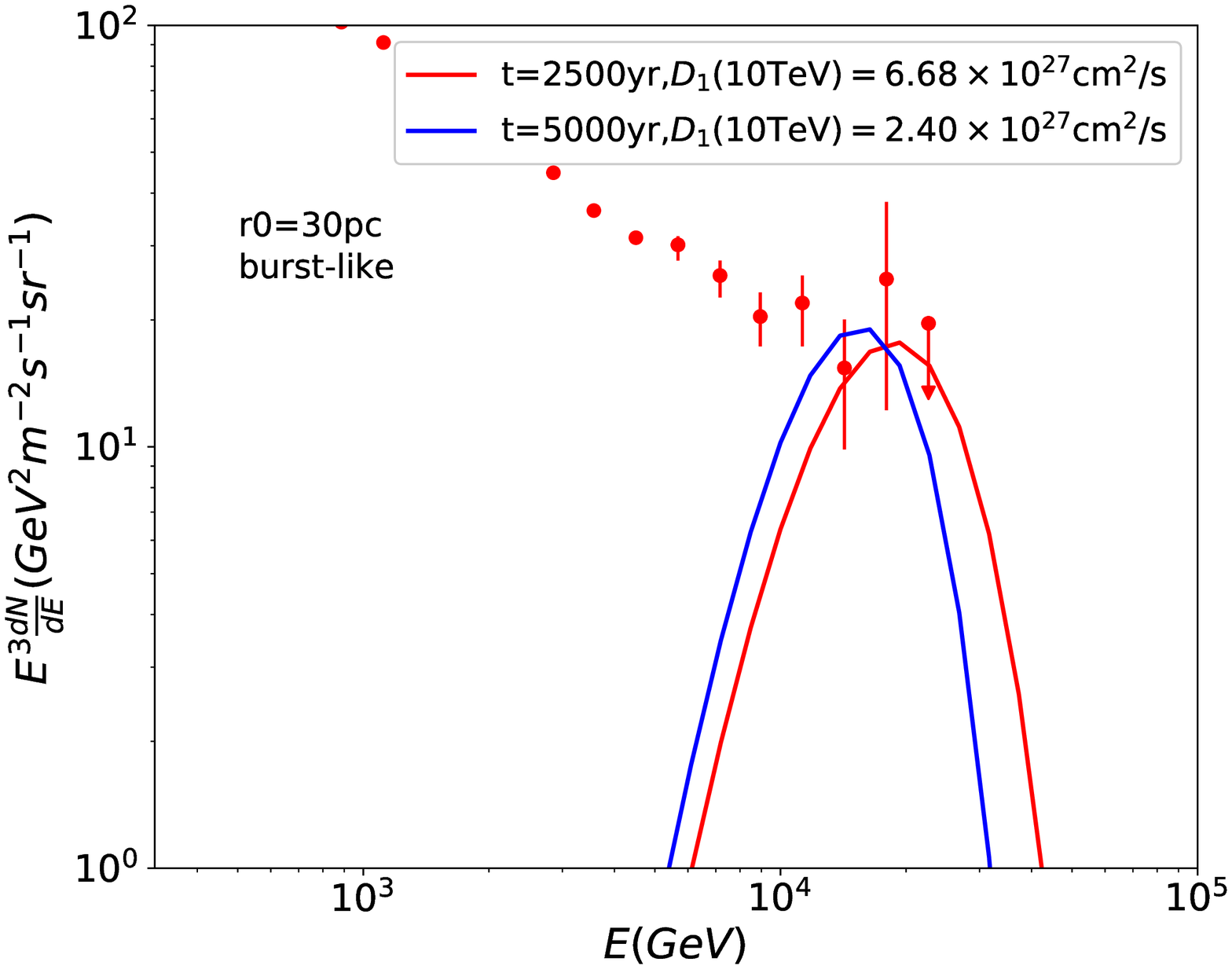}
\includegraphics[scale=0.34]{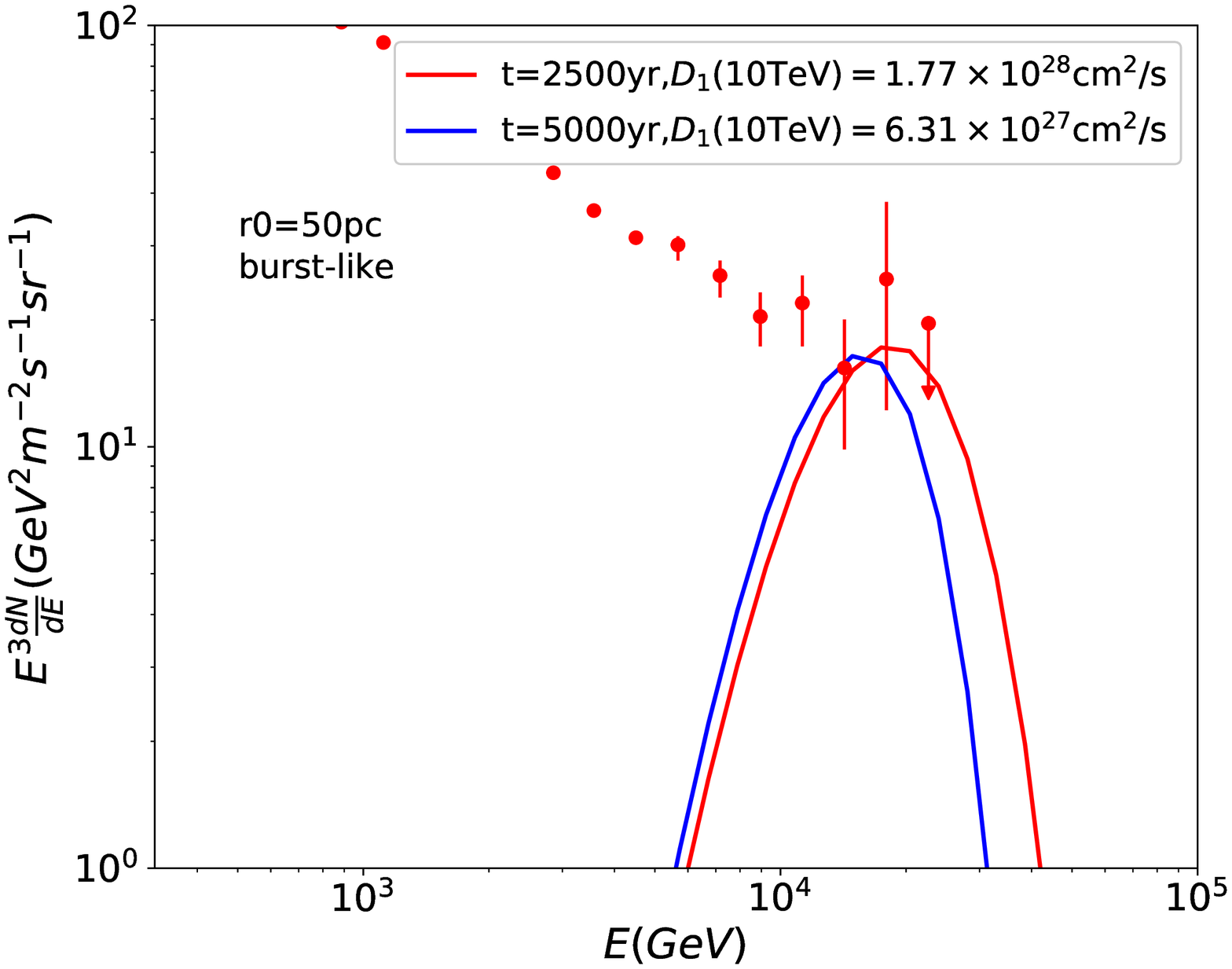}
\caption{Comparison between the CR electron spectrum produced by  Vela X in the two-zone model with different parameters. The burst-like injection is considered. We choose two reference values for the time $t$  when the burst-like injection occurs. Left panel: $r_0=$ 30 pc. Right panel: $r_0=$ 50 pc.}
\end{center}
\end{figure*}

In the above calculation, we have assumed $D(E)\propto E^{0.33}$. As the physics of the diffusion is not well-known, we also consider another case of the energy dependence, i.e., $D(E)\propto E^{0.548}$ for both the ISM and the region surrounding the PWN \citep{2017ApJ...836..172F}.  The results are shown in the left and right panels of Fig. 5 for  $r_0=30$ pc and $r_0=50$ pc respectively. We find that, for this kind of diffusion, we also have $D_1({\rm 10 TeV})\la10^{28}{\rm cm^2 s^{-1}}$, which is almost the same as the $D(E)\propto E^{0.33}$ case.

\begin{figure*}
\begin{center}
\includegraphics[scale=0.34]{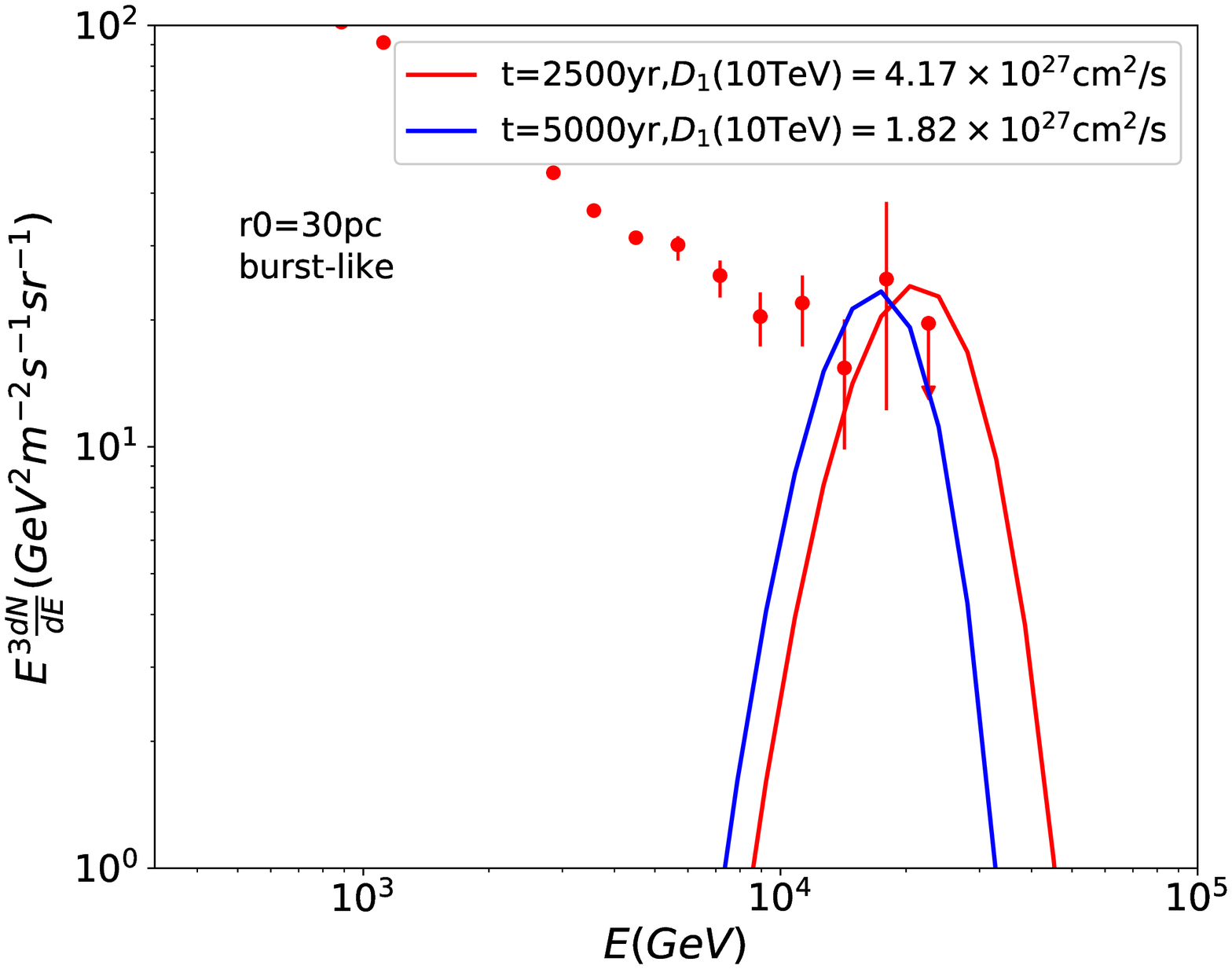}
\includegraphics[scale=0.34]{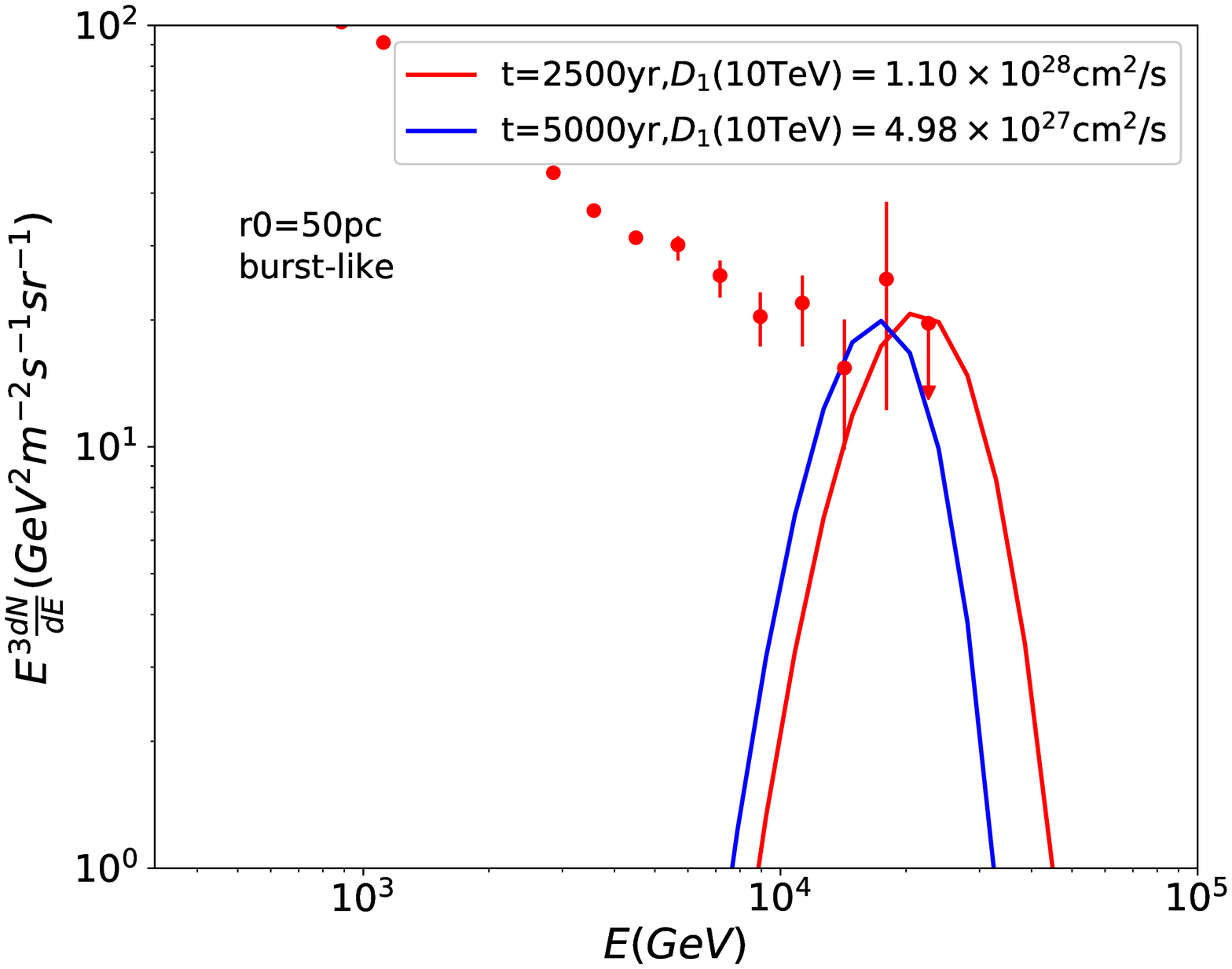}
\caption{The same as  Fig.4, but for the case where the diffusion coefficients in both inner and outer zones are assumed to be proportional to $E^{0.548}$ ($D_2(E) = 9.92\times 10^{27} (E/{\rm  GeV})^{0.548} \rm{cm^{2}/s}$) \citep{2018ApJ...854...57F}. Left panel: $r_0=$ 30 pc. Right panel: $r_0=$ 50 pc.}
\end{center}
\end{figure*}

{{In the above, we have taken a total injection energy of $6 \times 10^{48}$, as suggested by Hinton et al. (2011). To be more  conservative and  motivated by HAWC observations of other TeV halos, we also take a total injection energy corresponding to $\sim$ 10 percent of the total energy budget, i.e. $1.9 \times 10^{48}$ erg. The results are shown in Fig.6. It can be seen that the required $D_{1}$ does not change significantly, as the electron/positron flux is more sensitive to $D_{1}$ than the normalization of the injection spectrum.}}

\begin{figure*}
\begin{center}
\includegraphics[scale=0.34]{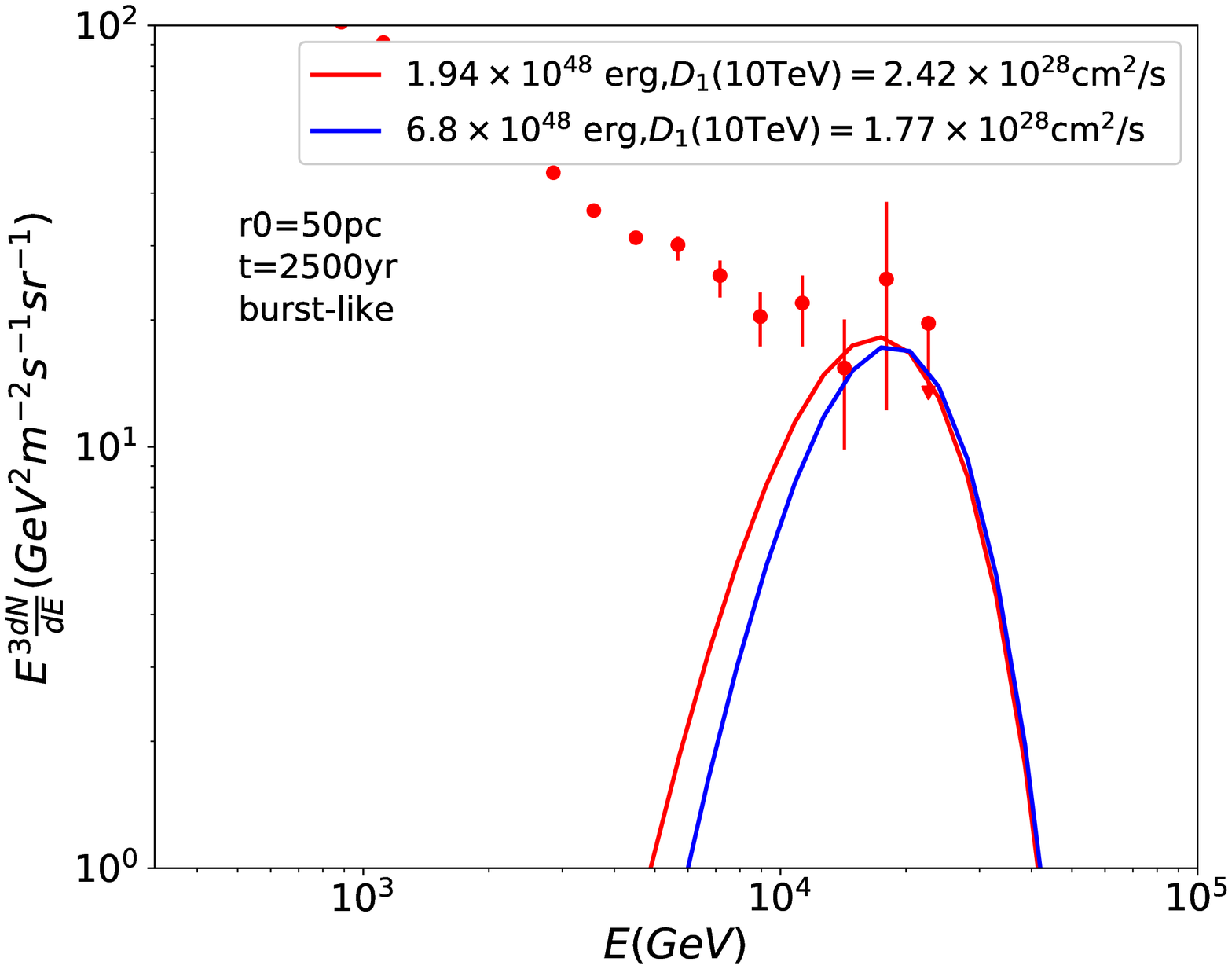}
\includegraphics[scale=0.34]{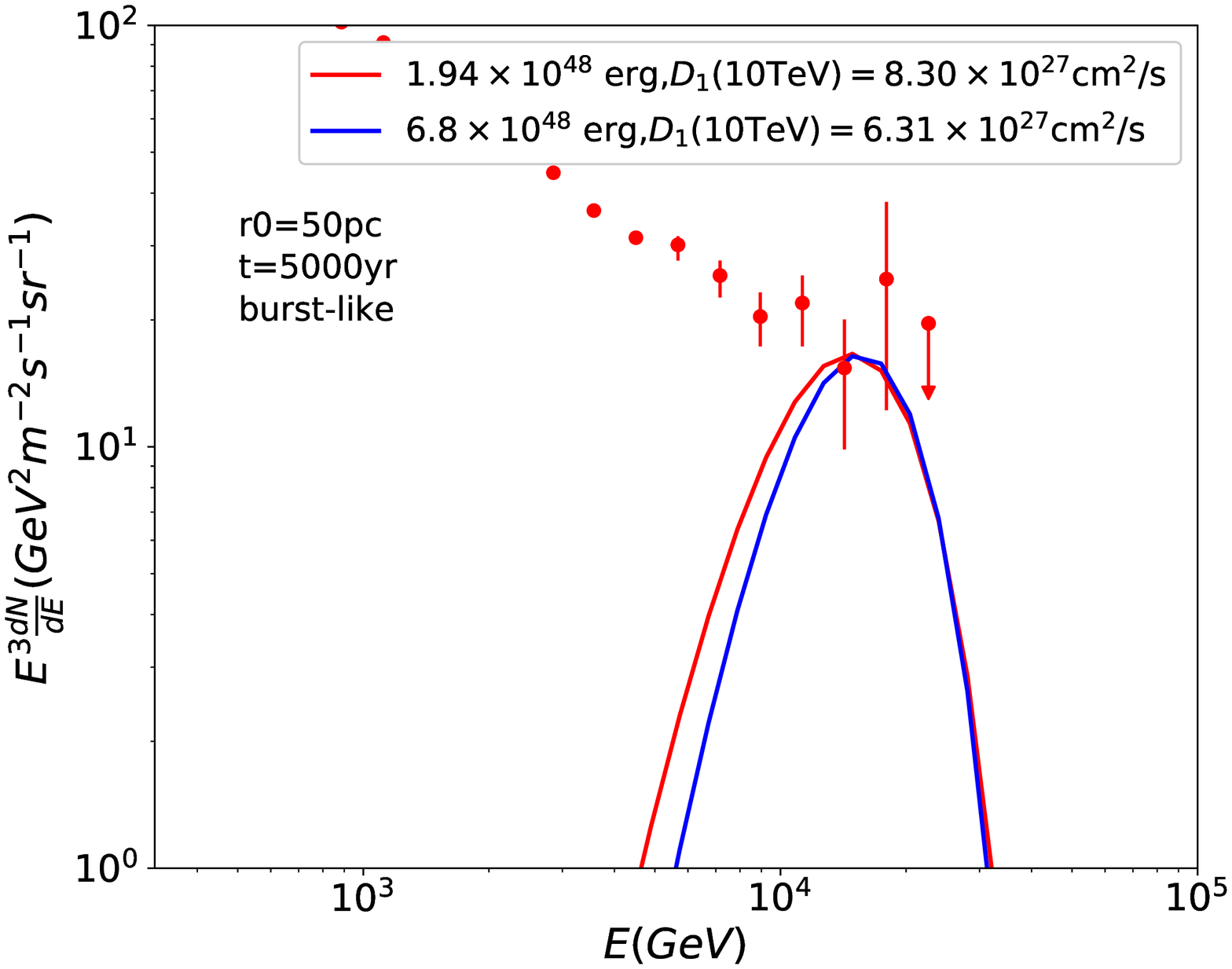}
\caption{The same as Fig.4, while different total electron energies are adopted. $r_0$=50 pc is adopted. The diffusion coefficients in both inner and outer zones are assumed to be proportional to $E^{0.33}$. Left panel: $t$= 2500 yr. Right panel: $t$= 5000 yr.}
\end{center}
\end{figure*}

{{The cutoff energy of the electron spectrum has been taken to be $E_{c}=6$ TeV following Hinton et al. (2011). We note that the spectra observed from Geminga and B0656+14 by HAWC favor much larger values of $E_{c}$, i.e., several tens of TeV. Results with different $E_{c}$ are shown in Fig.7. For a larger $E_{c}$, more electrons with higher energy are injected. As these energetic electrons can more easily escape from the slow-diffusion region and reach the Earth, the received flux of the higher energy electrons are significantly larger.  Therefore, the constraint on the diffusion coefficient becomes even more stringent and a smaller $D_{1}$ is required to reconcile with the HESS data.}}

\begin{figure}
\begin{center}
\includegraphics[scale=0.34]{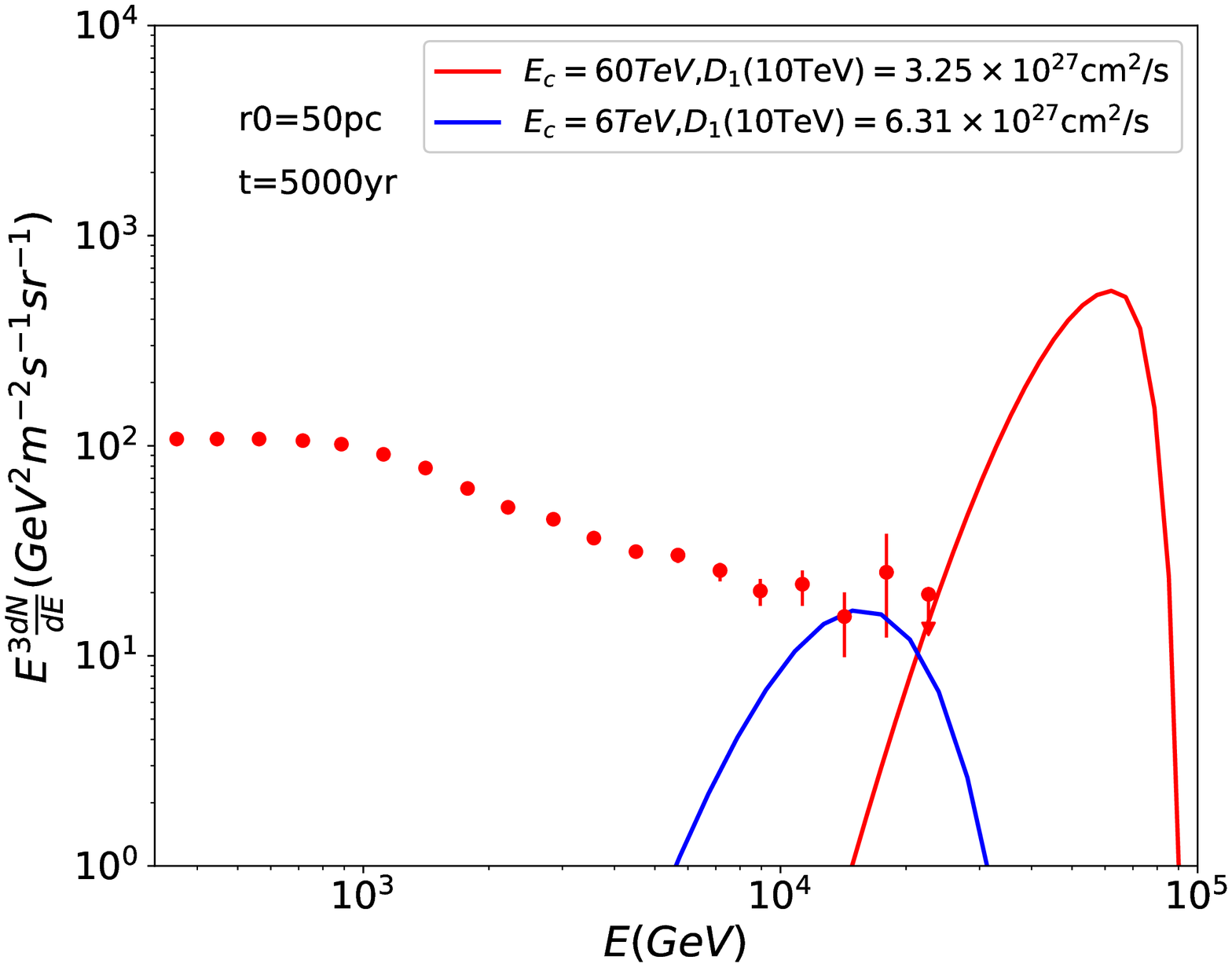}
\caption{Comparison between the CR electron spectrum with different $E_{c}$. The total injection energy is $6.8 \times 10^{48}$ erg. The diffusion coefficients in both inner and outer zones are assumed to be proportional to $E^{0.33}$. The red and the blue lines represent the injection spectra with $E_c =$ 60 TeV and 6 TeV, respectively.}
\end{center}
\end{figure}

{{In all of these cases, $D_1({\rm 10 TeV})\la10^{28}{\rm cm^2 s^{-1}}$ can be obtained. Thus,  the limit on the diffusion coefficient is robust, independent of  uncertainties in the input parameters.  We conclude that the diffusion in the immediate vicinity of  Vela X must be highly inefficient.}}

\section{Summary}
There have been suggestions that electrons must have undergone diffusive escape from the the Vela X PWN,  indicated by the evidence for a roll-over of the electron spectrum at energies of a few tens of GeV. These escaped electrons may contribute to the CR electron flux at the earth.
In this paper, we have shown that recent HESS data of the CR electron flux at $\sim 10$ TeV place interesting constraints on the diffusion coefficient around Vela X.
We find that a highly inefficient diffusion region in the immediate vicinity of Vela X must be present, with $D({\rm 10 TeV})\la10^{28}{\rm cm^2 s^{-1}}$ . The result is consistent with the recent finding that there are inefficient diffusion regions surrounding the Geminga  and PSR B0656+14 PWNe, suggesting that such inefficient diffusion regions may be common around PWNe with various ages.

Previous theoretical studies have suggested that CRs  can be scattered by the self-generated Alfv{\'e}n waves induced by streaming instability \citep{2004ApJ...614..757Y,2012ApJ...745..140Y,2013ApJ...768...73M,2018arXiv180701608Q}, whereby the particles are self-confined. If the CR flux is sufficiently high, the growth rate of the streaming instability can dominate the nonlinear damping of background turbulence. We speculate that such a process enhances the CR scattering rate inside $r_0$ and hence reduces the diffusion coefficient to the required level. Another possible mechanism is the influences of the Vela SNR surrounding Vela X. The turbulence in the areas swept up by the shock of the SNR can be strong, which may induce a smaller diffusion coefficient \citep{1978MNRAS.182..147B}.

\acknowledgements We thank Dan Hooper and Huirong Yan for helpful discussions.  X.Y. Wang is supported by the National Key R \& D program of China under the grant 2018YFA0404200, 973 program under grant 2014CB845800, the NSFC under grants 11625312 and 11851304.

\newpage

\appendix
We first compare the CR flux obtained with our analytic approach for the two-zone model with $D_1=D_2$ and the simple one-zone model with the same diffusion coefficient (i.e., $D=D_1$). The result is shown in Fig.8. One can see that the difference between the two models is negligibly small. This demonstrates that the treatment
described by Fig.2 is quite accurate.

\begin{figure}
\begin{center}
\includegraphics[scale=0.3]{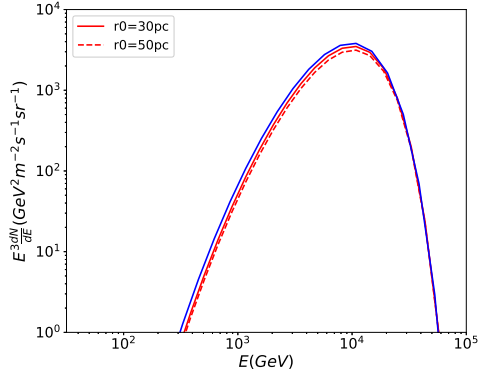}
\caption{Comparison between the CR spectrum obtained with our analytic approach for the two-zone model with  spatially-independent  diffusion coefficient  (i.e. $D_1 =D_2 =3.86 \times 10^{28} (E_e/\rm{GeV})^{0.33} \rm{cm^{2}/s}$) and that obtained with the simple one-zone model (i.e., Eq. 4) with $D=3.86 \times 10^{28} (E_e/\rm{GeV})^{0.33} \rm{cm^{2}/s}$. A burst-like injection is considered. The blue solid line represents the result obtained using Eq.4, while the red solid line and the red dashed line represent the results obtained for the two-zone analytic model  with $r_0=$30 pc and 50 pc, respectively. The injection time is taken to be $t=2$ kyr.}
\end{center}
\end{figure}

However, this test does not account for the influence caused  by the difference in the diffusion coefficients of the two zones (i.e., $D_1\neq D_2$).
There are two factors that may cause differences between the results in our analytical model and that in the realistic case. The density gradient of electrons at $r_0$ calculated by Eq.4 becomes smaller when $D_2>D_1$, since a larger diffusion coefficient in the outer zone leads to a faster diffusion outward. Thus, our model underestimates the flux escaping from the spherical surface at $r_0$ in this regard.
On the other hand, part of the electrons located at the spherical surface at $r_0$  will go back to the inefficient diffusion zone. These electrons will take  longer time to arrive at the earth. Our analytic estimate does not take into account this effect, so our   result overestimate the CR flux at the earth in this regard.

To test the precision for the case of $D_1\neq D_2$, we compare our analytic result with the numerical result obtained by Fang et al. (2018), which solves  Eq.1 for the case of $D_1\neq D_2$ with a numerical method. The results are shown in Fig.9. For the cases of $t=2500{\rm yr}$, the difference between our analytic result and that obtained with the numerical method is at most $30\%$.
For the cases of $t= 5000 \rm{yr}$, the difference is even smaller.

\begin{figure*}
\begin{center}
\includegraphics[scale=0.34]{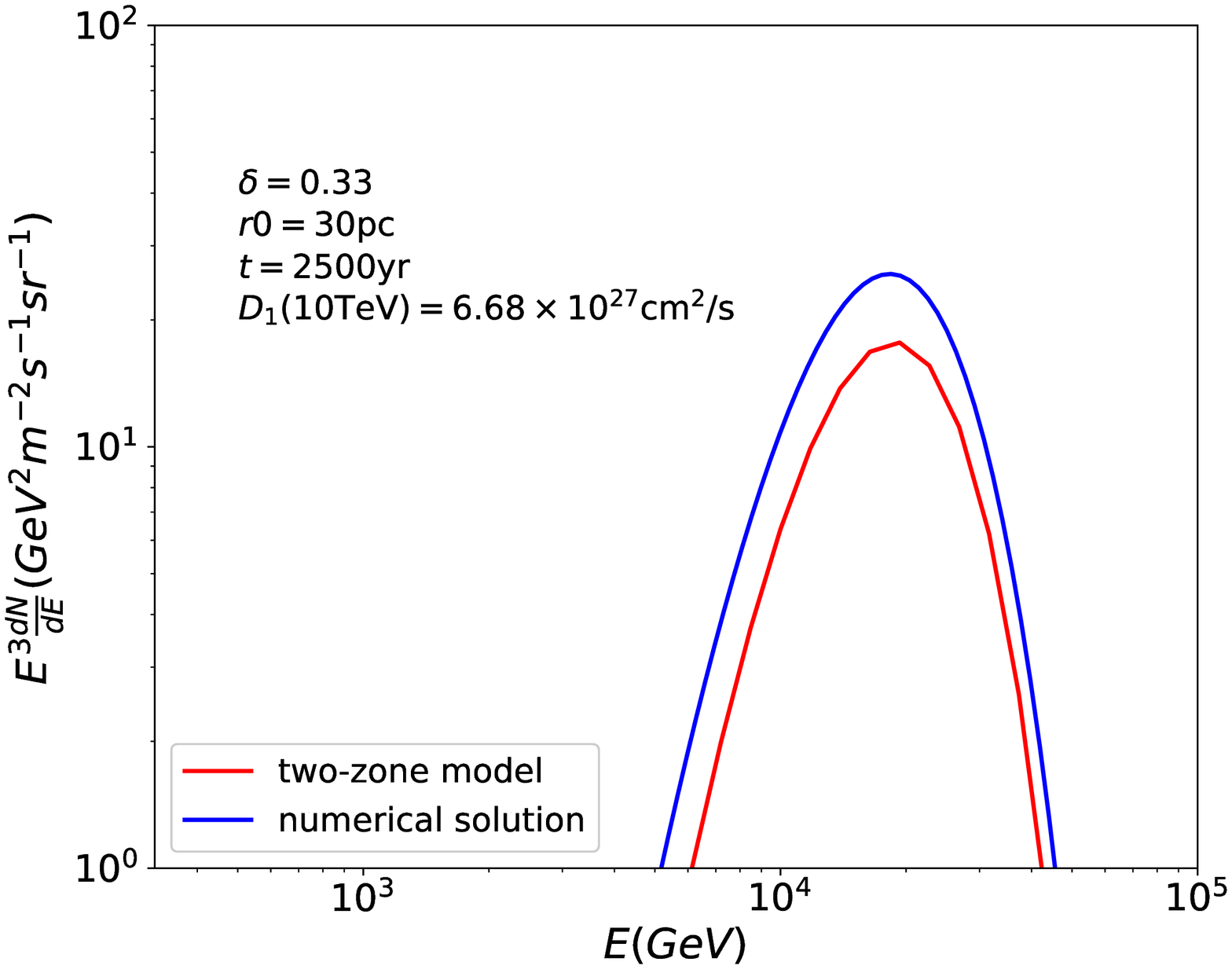}
\includegraphics[scale=0.34]{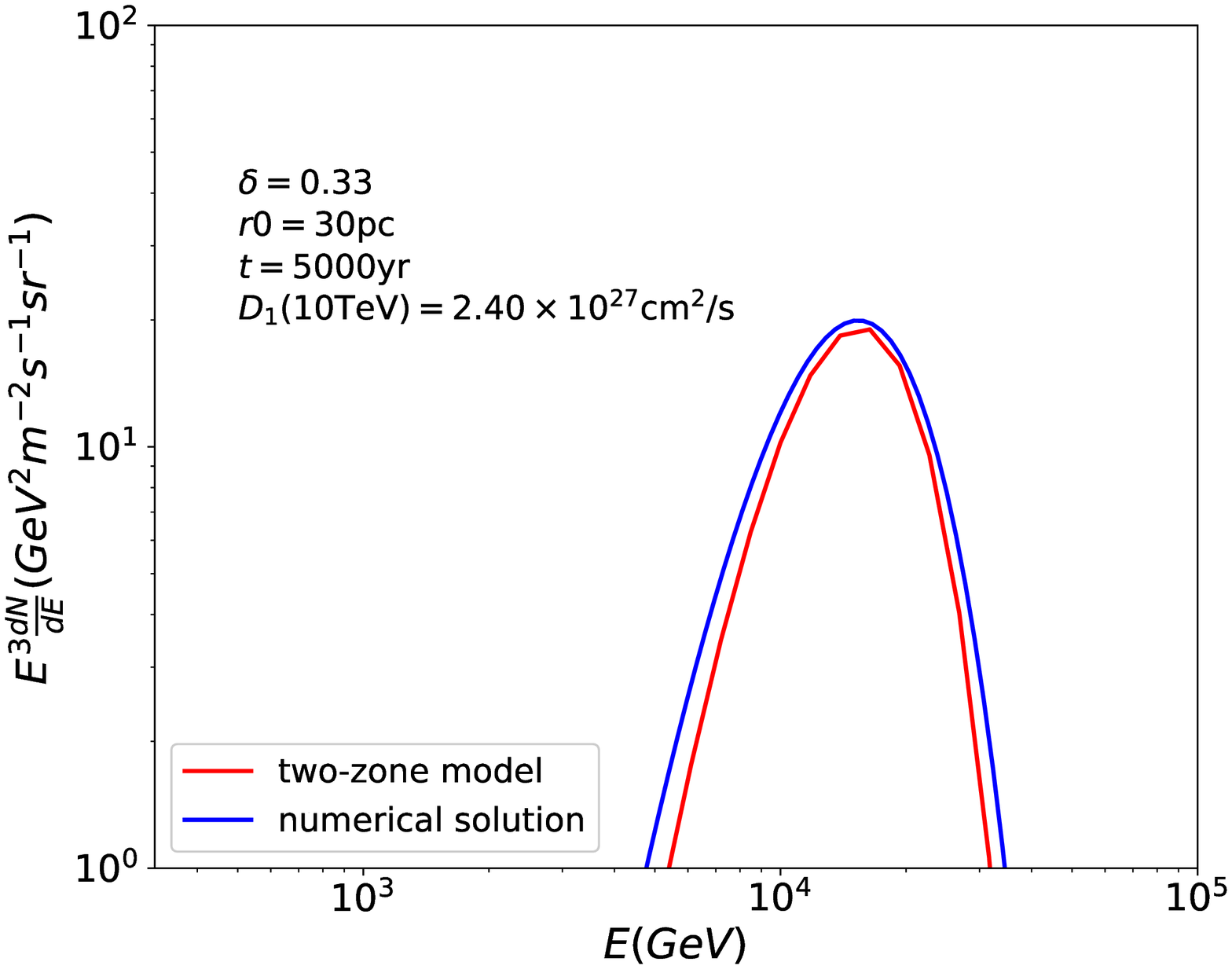}
\includegraphics[scale=0.34]{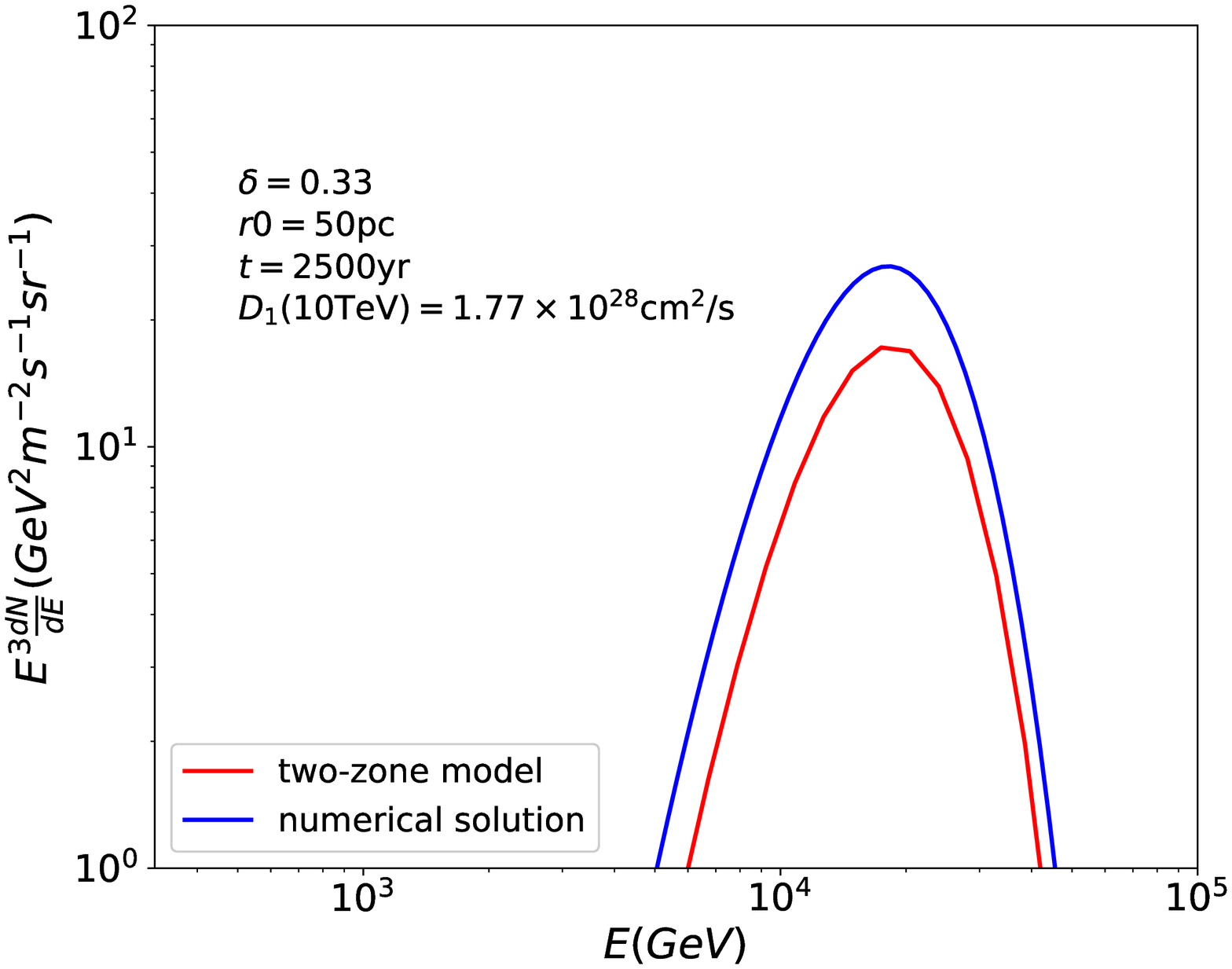}
\includegraphics[scale=0.34]{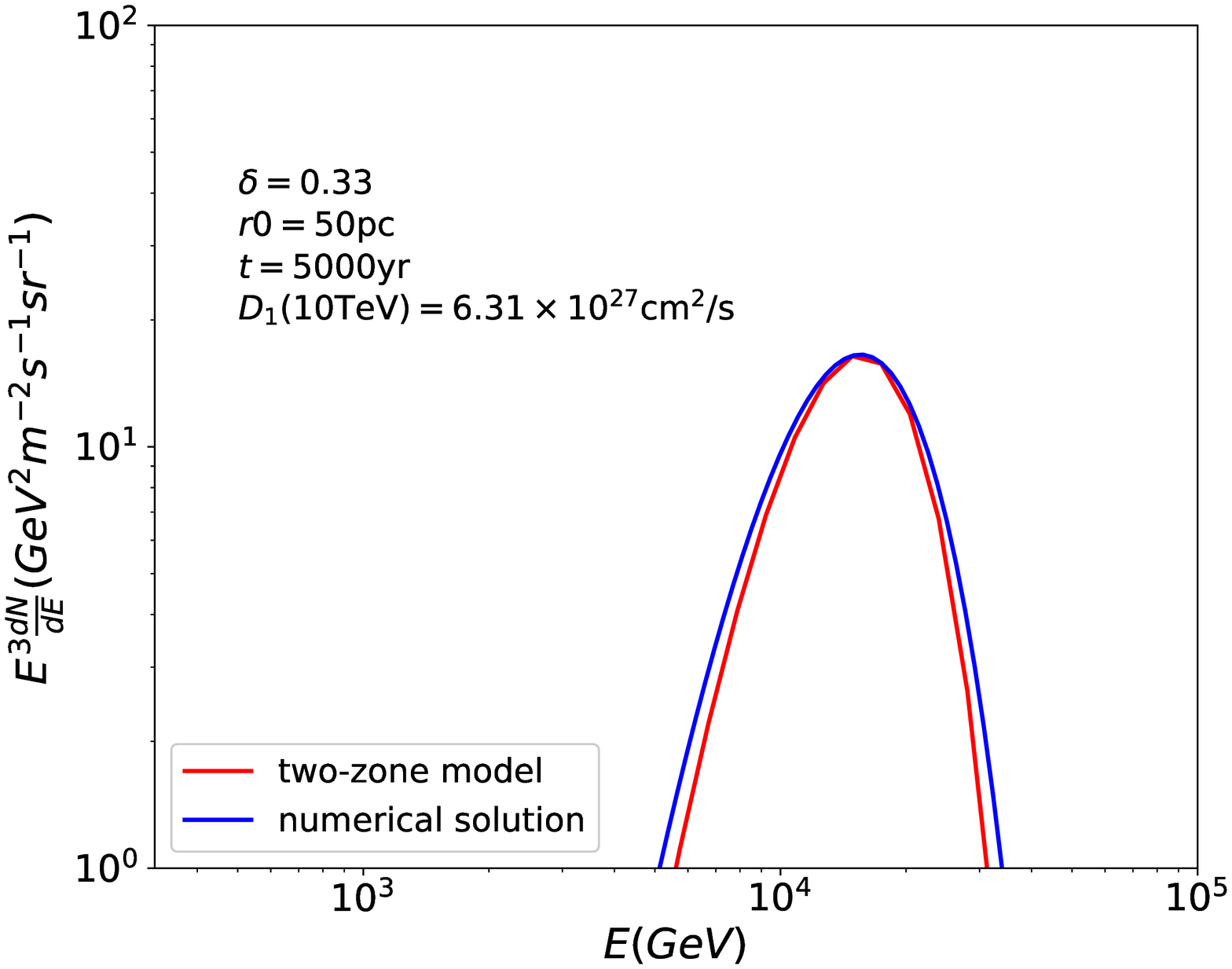}

\caption{Comparison between the results in our analytic two-zone model and the numerical solution in \cite{2018arXiv180302640F}. A burst-like injection is adopted. Four groups of parameters are tested. The red lines represent the flux in our analytic model and the blue lines represent the results in the numerical solution.}
\end{center}
\end{figure*}

\end{document}